\documentclass[
 aip,
 amsmath,amssymb,
reprint,%
]{revtex4-1}

\usepackage{graphicx}
\usepackage{dcolumn}
\usepackage{bm}

\usepackage[utf8]{inputenc}
\usepackage[T1]{fontenc}
\usepackage{mathptmx}

\usepackage{xcolor}
\usepackage{hyperref}

\usepackage[toc,page,titletoc]{appendix}
\usepackage[capitalise,nameinlink]{cleveref}

\usepackage[rightcaption]{sidecap}
\usepackage{setspace}

\newcounter{line}[part]

\newcounter{algo}[part]

\begin{document}

\preprint{AIP/123-QED}

\title[]{Algorithm for automated tuning of a quantum dot into the single-electron regime}

\author{M.~Lapointe-Major}%
\affiliation{Institut quantique and Département de Physique, Université de Sherbrooke, Sherbrooke, Québec, J1K 2R1, Canada}%

\author{O.~Germain}%
\affiliation{Département de Mathématiques, Université de Sherbrooke, Sherbrooke, Québec, J1K 2R1, Canada}%

\author{J.~Camirand~Lemyre}%
\affiliation{Institut quantique and Département de Physique, Université de Sherbrooke, Sherbrooke, Québec, J1K 2R1, Canada}%

\author{D.~Lachance-Quirion}
\affiliation{Institut quantique and Département de Physique, Université de Sherbrooke, Sherbrooke, Québec, J1K 2R1, Canada}%

\author{S.~Rochette}
\affiliation{Institut quantique and Département de Physique, Université de Sherbrooke, Sherbrooke, Québec, J1K 2R1, Canada}%

\author{F.~Camirand~Lemyre}%
\affiliation{Département de Mathématiques, Université de Sherbrooke, Sherbrooke, Québec, J1K 2R1, Canada}%

\author{M.~Pioro-Ladrière}
\affiliation{Institut quantique and Département de Physique, Université de Sherbrooke, Sherbrooke, Québec, J1K 2R1, Canada}%

\date{\today}


\begin{abstract}
We report an algorithm designed to perform computer-automated tuning of a single quantum dot with a charge sensor.  The algorithm performs an adaptive measurement sequence of sub-sized stability diagrams until the single-electron regime is identified and reached.  For each measurement, the signal processing module removes the physical background of the charge sensor to generate a binary image of charge transitions.  Then, the image analysis module identifies the position and number of lines using two line detection schemes that are robust to noise and missing data.
\end{abstract}

\maketitle

Spin qubits in quantum dots are among the frontrunner architectures for the implementation of a small-scale quantum computer \cite{Promissing_a, Promissing_b} due to their high potential for scalability \cite{Scalability_a, Scalability_b, Scalability_c, SplitGate}.  Indeed, progress towards devices with multiple quantum dots has recently been demonstrated \cite{NineDotArray, ElectronShuttling}. However, as the number of quantum dots increases, the brute-force approach of manually adjusting several gate voltages per quantum dot to reach the qubit regime has become impractical.  To date, softwares have been developed to address this issue by automatizing tedious parts of this process for double dots using image analysis or machine learning tools to adjust the inter-dot tunnel coupling \cite{Automated_InterdotCoupling}, detect triple points in stability diagrams \cite{AutomatedTuningDelft, TriplePointAndrew} and perform state recognition \cite{AutomatedTuningMachineLearning_a, AutomatedTuningMachineLearning_b}.


In this paper, we report an algorithm designed to perform automated tuning of a single quantum dot tunnel-coupled to a reservoir of electrons using only charge sensing.  This has been recognized as challenging for the following reasons: (i)~tuning a single dot requires line detection, which proves to be less robust than the detection of triple points \cite{AutomatedTuningDelft}; (ii) the number of transitions in a measurement is \textit{a priori} unknown; (iii) the detection of transition lines with possible curvature and in the presence of noise and missing data points is computationnally expensive \cite{ReviewHough}; (iv) the charge sensor couples to all charges at proximity, thus measuring several unwanted features giving rise to a physical background in the resulting signal.



\begin{figure}[ht]
\includegraphics{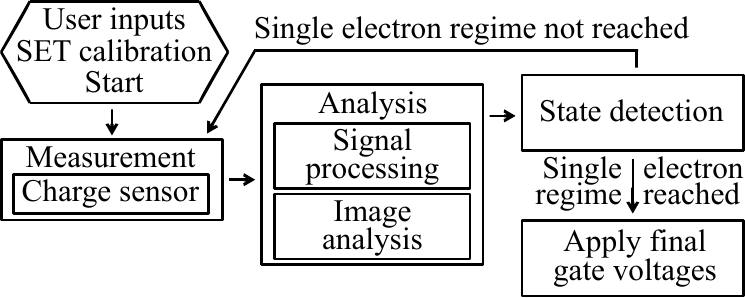}
\caption{\label{fig_flowchart} Flow chart of the algorithm.  A stability diagram is first measured by the charge sensor.  This measurement goes through the signal processing module, which removes the physical background of the charge sensor and generates a binary image of the detected transitions.  The image analysis module then identifies transition lines in that binary image.  If the single-electron regime is reached, the algorithm sets gate voltages appropriately, and otherwise loops again in the measurement/analysis sequence.}
\end{figure}

Our algorithm achieves all this by performing an iterative sequence of measurements, analysis and state detection until the single-electron regime is reached (Fig. \ref{fig_flowchart}) and only requires the charge sensor to be pre-calibrated with minimal user inputs.  Given a measurement, the signal processing module removes the physical background of the charge sensor to generate a map of detected charge transitions and the image analysis module reconstructs the transition lines from that map.  The latter was implemented using two different approaches, namely a modified Hough transform \cite{HoughImproved} and EDLines algorithm \cite{EDLines}, each of which have different advantages regarding computation time and detection of curved transitions.  Finally, a sequence of measurements that allows the algorithm to reach the single-electron regime is introduced.




\begin{SCfigure*}
\includegraphics{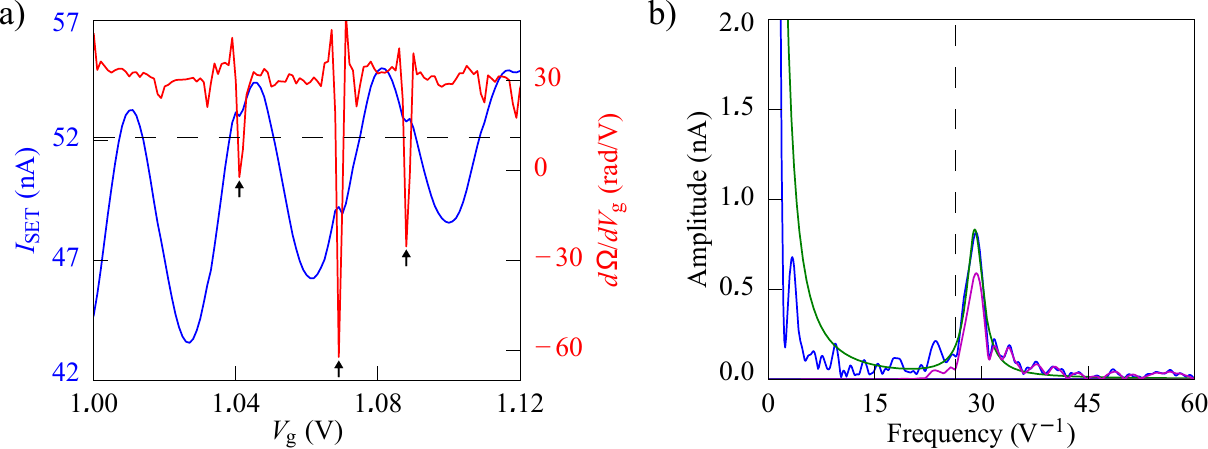}
\caption{\label{sigpro} \textbf{(a)}~ Typical $I_\mathrm{SET}$ trace (blue) and instantaneous frequency of the $I_\mathrm{SET}$ oscillations (red) as a function of gate voltage $V_\mathrm{g}$.  Arrows indicate detected transitions and the dashed line the threshold computed for this frequency distribution.  \textbf{(b)}~ Fourier transform of the data shown in (a) (blue), Lorentzian fit (green) used to extract the cutoff frequency for the high-pass filter (vertical dashed line) and the Fourier transform of the filtered signal (purple).}
\end{SCfigure*}

In our setup, a single quantum dot is tunnel-coupled to a reservoir of electrons in the split accumulation gate geometry identical to the device in Fig. 1 of Ref \cite{SplitGate}.  The quantum dot is capacitively coupled to a single electron transistor (SET), used for charge sensing.  The current through the SET ($I_\mathrm{SET}$) is measured with a 1 MHz bandwidth cryogenic amplifier \cite{CryoAmp}.  In that setup, $I_\mathrm{SET}$ is monitored as a function of gate voltages (Fig. \ref{sigpro}a) to detect transitions in the quantum dot electron occupancy.  We model the current through the SET following equations (\ref{eq_ISET}) and (\ref{eq_SinArg}).

\begin{subequations}
\begin{align}
\label{eq_ISET}
I_\mathrm{SET} &= A(V_\mathrm{g}) \sin [\Omega (V_\mathrm{g}) ] + B(V_\mathrm{g}),\\
\label{eq_SinArg}
\Omega (V_\mathrm{g}) &= \omega (V_\mathrm{g})\cdot V_\mathrm{g}+\phi(N),
\end{align}
\end{subequations}

The current contains a zero and low-frequency background term $B(V_\mathrm{g})$ and an oscillating term with a voltage-dependent amplitude $A(V_\mathrm{g})$ and argument $\Omega (V_\mathrm{g})$.  In general, the frequency $\omega (V_\mathrm{g})$ can depend on the gate voltage, since the charging energy of the SET can vary over large ranges of gate voltage \cite{ChargingEnergy}.  Here, this effect is mitigated by performing measurements over ranges such that this effect can be neglected, leading to $\omega(V_\mathrm{g})\approx\omega$.  The term $\phi(N)$ models jumps that occur when an electron is added in the quantum dot.  Because the transitions occur for specific voltages, $\Omega(V_\mathrm{g})$ can be rewritten as $\Omega(V_\mathrm{g}) = \omega (\Delta N(V_\mathrm{g}))\cdot V_\mathrm{g}$, where $\Delta N$ denotes transitions of $N$ and $\omega$ is constant except at specific voltages where a charge transition occurs, for which $\omega$ becomes highly negative.  For the purpose of charge detection, the only components of interest in $I_\mathrm{SET}$ are the transitions in the electron occupancy of the quantum dot, $\Delta N$.  

In the signal processing module, the goal is to identify gate voltages for which a charge transition $\Delta N$ occurs.  This is achieved by removing all the other components in the signal.  First, a fifth-order Butterworth high-pass filter \cite{ButterworthBook} is applied on the measured signal to remove the background term $B(V_\mathrm{g})$.  The cutoff frequency for the filter is extracted by fitting a Lorentzian to the Fourier transform of the signal (Fig. \ref{sigpro}b).  Next, the instantaneous frequency of the oscillations is extracted using a Hilbert transform of the filtered signal (Fig.~\ref{sigpro}a).  The instantaneous frequency shows negative jumps at gate voltages where a charge transition occurs.  These jumps are identified by computing an adaptive threshold ($T$) taking into account the average ($\bar{\omega}$) and standard deviation ($\sigma_\omega$) of the distribution of frequencies.  Any point below the threshold is then identified as a charge transition.  Using a severe threshold ($T$>$\bar{\omega}$-3.5$\sigma_\omega$) yields several false-negative results while a loose threshold ($T$<$\bar{\omega}$-2$\sigma_\omega$) yields several false-positive results.  We find the image analysis module to perform best using $T$=$\bar{\omega}$-3$\sigma_\omega$ for threshold.

\begin{SCfigure*}
\includegraphics{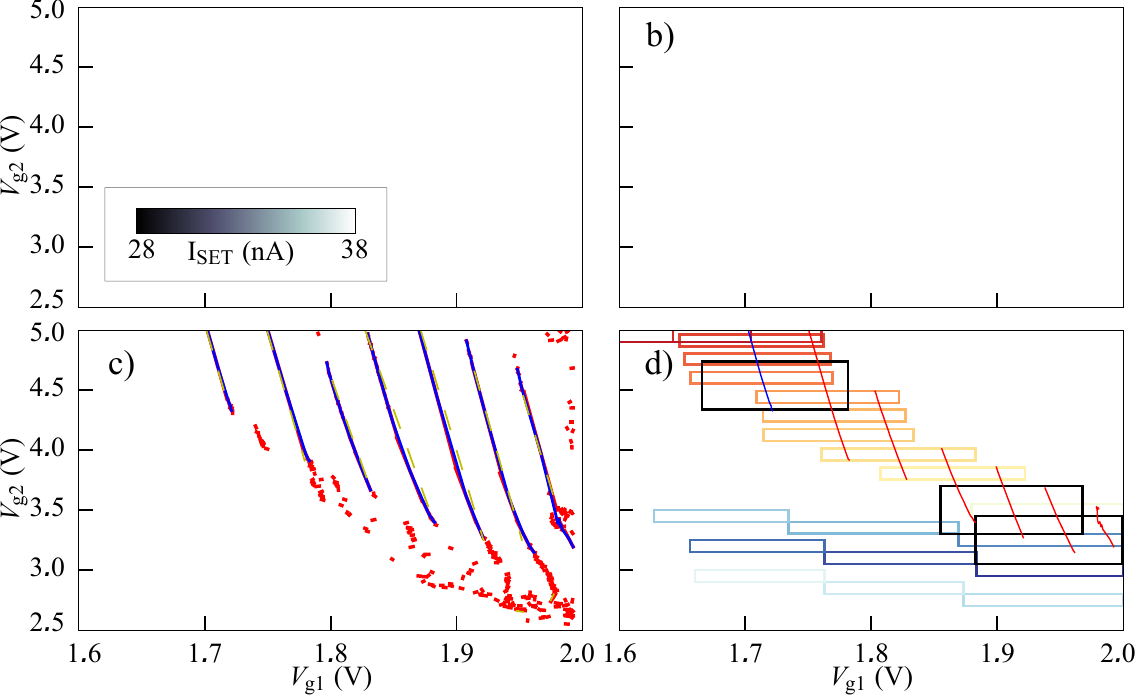}
\caption{\label{diag} \textbf{(a)}~ Experimentally measured stability diagram on a split accumulation gate quantum dot.  $V_\mathrm{g1}$ is the dot gate with a 0.5 mV resolution (801 pixels) and $V_\mathrm{g2}$ is the reservoir gate with a 2.5 mV resolution (1001 pixels).   \textbf{(b)}~ Binary image of the transition points detected by the signal processing module for the stability diagram shown in (a).  \textbf{(c)}~ Transition lines detected by the image analysis module with the EDLines algorithm (yellow) or the modified Hough transform before (red) and after (blue) the line reconstruction protocol.  \textbf{(d)}~ Measurement sequence used by the algorithm to reach the single-electron regime.  The first sub-diagram sampled is in dark blue and the last in dark red.  Larger sub-diagram sampled to confirm transitions are in black.  Transition lines detected by the image analysis after the tuning process is completed are shown in red and the first charge transition is in blue.}
\end{SCfigure*}

Typical measurements used to identify charge transitions are two-dimensional stability diagrams where the voltage of a first gate ($V_\mathrm{g1}$) is swept and the voltage of a second gate ($V_\mathrm{g2}$) is stepped after every $V_\mathrm{g1}$ sweep (Fig.~\ref{diag}a).  Two-dimensional measurements provide many advantages to the algorithm: (i)~it improves robustness to noise when false-positive transitions are detected; (ii) it allows the algorithm to extrapolate information in regions where the charge sensor sensitivity is reduced and (iii) it provides information about the ratio of the lever arms of the two gates.  

Even though the measured stability diagrams are two-dimensional, the signal processing procedure is kept one-dimensional to circumvent the detection of telegraphic noise as a transition line by the image analysis module.  The result of the signal processing module applied on a large stability diagram is a binary map of detected transitions (Fig. \ref{diag}b).

Compared to other existing solutions to remove the physical background from charge sensor measurements \cite{SETCompensation, RF_Reflecto}, this approach does not require additional hardware and only requires post-measurement processing.



Following the signal processing module, the image analysis module is used to identify transition lines in a measurement.  The main challenge of this step arises from the charge sensor not providing any way to label the detected transition points.  This means the algorithm must identify (i)~to which transition line each of the transition points belongs to, (ii) the number of transition lines in a given measurement and (iii) their position.  Measured transitions must be detected amidst device and signal processing noise, missing data points and curvature of the transitions.  This has led to two implementations for this module using either a modified Hough transform \cite{HoughImproved} or EDLines \cite{EDLines} for the efficient detection of curved lines or fast computation time respectively.


The modified Hough transform is implemented in a divide-and-conquer approach, where segments are recursively split into smaller, more manageable ones before being either discarded or reconstructed into charge transition lines.  First, the algorithm generates clusters of points using a linkage algorithm.  Each cluster is defined to be a distance of at least one pixel from its neighbouring clusters.  For each of the clusters generated this way, a modified Hough transform is computed which estimates the best segment describing the cluster along with the covariance matrix as described in Ref. \cite{HoughImproved}.  Here, a correction is applied to the covariance matrix to account for the width of the transition lines in the stability diagrams, which comes from the statistical nature of the tunnelling events and the tunnel coupling between the dot and the reservoir.  Following Ref. \cite{ThreeDRecognition}, clusters are recursively split to break down curved clusters into smaller pieces that can be appropriately approximated by a linear segment and transition lines are reconstructed based on collinearity and proximity of the endpoints criteria (Fig. \ref{diag}c).  



The second implementation uses the EDLines algorithm in a top-down approach.  The EDLines alorithm identifies anchor points to generate segments with very few false-positives.  Given the width of charge transitions, the EDLines algorithm detects several segments per transition.  Doublons are discarded using the parallelism criteria in Ref.~\cite{ThreeDRecognition}.  Segments are regrouped and identified as transition lines based on the same collinearity and proximity of the endpoints criteria used for the modified Hough transform.



Both implementations of the image analysis module are robust to missing data points (false-negatives) and noise (false-positives) \cite{HoughImproved, EDLines}.  This has been verified on a number of stability diagrams acquired on two different devices and is pictured in Fig. \ref{diag}c where line reconstruction succeeds despite noise surrounding the transitions.  On these devices, false-positive and false-negative instances are always in sufficiently low occurence to guarantee success of our algorithm.  Missing data can occur for a variety of reasons: (i) due to a loss of sensitivity of the SET, which can be caused by the modulation of the tunnel barriers in the SET due to interface irregularities \cite{ModTunnelBarriers}~; (ii) it can be introduced by the signal processing module when using a severe threshold $T$ or (iii) introduced by the measurement sequence (Fig. \ref{diag}d).  Typical sources of noise in stability diagrams are charge noise, current noise in the SET, telegraphic noise and false-positive transition points added by the signal processing module.  The similar performances of the two algorithms in the presence of typical noise is explained by the fact that EDLines identifies very few false-positive transition lines by design \cite{EDLines} while the line reconstruction of the modified Hough transform is very efficient at identifying and discarding them (red segments in Fig. \ref{diag}c).


The modified Hough transform implementation is expected to be more efficient at detecting charge transitions with large curvature because of the initial construction of transition lines, which regroups all transition points into a cluster independent of the curve before splitting and reconstructing it in its divide-and-conquer scheme.  These transitions with large curvature would not be detected by the EDLines algorithm due to its validation method \cite{EDLines}.  This has been qualitatively observed in some experimental datasets.  For the stability diagram in Fig. \ref{diag}c, the image analysis requires 45 seconds to identify the blue transition lines using the modified Hough transform implementation and requires 1.5 seconds to identify the green dashed transition lines using the EDLines implementation on an Intel(R) Xeon(R) CPU E3-1245 v5 running at 3.50 GHz. This speed-up is consistent with results obtained on other experimental stability diagrams. 

Following the flow chart of Fig. \ref{fig_flowchart}, a heuristic algorithm is now applied to find the last charge transition.  The goal of the measurement sequence is to acquire enough information about the quantum dot to identify the single-electron regime with the fewest measurements possible.  This heuristic mimics a typical tuning protocol, where the dot is first formed and emptied until the last transition is found based on sub-sized stability diagrams measured using only the two accumulation gates ($V_\mathrm{g1}$ and $V_\mathrm{g2}$).  This is reasonable since these two gates offer full control over physical quantities meaningful to the algorithm in the split accumulation gate geometry \cite{SplitGate}.  


A sub-diagram is first measured in a suitable user-specified scan range.  If no transition is detected, the algorithm performs diagonal series of measurements until a charge transition is detected.  To verify the detected line truly belongs to a transition and is not an artifact of the signal processing or due to experimental noise, a larger sub-diagram is measured centered on the detected line.  The program then follows the transition by increasing $V_\mathrm{g2}$ until it disappears due to broadening.  Then, lower $V_\mathrm{g1}$ measurements are performed until the next transition is found.  This is done recursively until no more transitions are found, meaning the quantum dot is empty.  Finally, the program analyzes the ensemble of all completed measurements and identifies the first charge transition (Fig. \ref{diag}d).  A larger sub-diagram is then measured on this transition to confirm it and gate voltages are adjusted to add one electron back into the quantum dot.  This measurement sequence was tested by sampling sub-diagrams from large experimental stability diagrams taken on two different devices.

The measurement sequence takes advantages of the robustness to missing data of the image analysis module by leaving blank spaces between measurements (Fig. \ref{diag}d), which allows to reduce measurement time.  Segments from different measurements are regrouped only at the end of the sequence.  This reduces the computational cost as the image analysis module is called only once per measurement with one extra call for the sum of all measurements.  

The scan range of each sub-diagram is limited by the signal processing module.  The extraction of the cutoff frequency for the high-pass filter and the Hilbert transform both yield significantly better results when a full cycle of the SET background oscillations is acquired.  Therefore, prior to the measurement of a sub-diagram, a preliminary sweep of $V_\mathrm{g1}$ is performed to estimate the background frequency.  The width of the stability diagram is then determined to include at least a full cycle of the oscillations.  The height of the stability diagram is arbitrarily chosen to be 40 pixels, which yields a good enough transition length for the line reconstruction given a typical transition width, which is three or four pixels wide when using a gate voltage resolution of approximately 1 mV.  The voltage resolution is user-specified and kept fixed through all measurements to avoid pixel connectivity issues that would arise due to changing pixel size and voltage grid.

In summary, we have developed an algorithm designed to tune a single quantum dot to the single-electron regime.  We have shown a protocol to remove the physical background from charge sensor measurements that can loosen the requirements for additional hardware and feedback loops in SET-based charge detection.  Furthermore, we have developed and compared two image analysis algorithms to identify charge transitions.  While EDLines is at least ten times faster, the modified Hough transform approach is believed to present additional robustness to line curvature.  We envision that combined with recently demonstrated identification of triple points \cite{AutomatedTuningDelft, TriplePointAndrew}, our algorithm will provide additional tools for automated initialization and control routines of quantum dots.

\section*{Supplementary material}

See supplementary material for pseudo-code summarizing key steps of the algorithm.

\begin{acknowledgments}
We recognize helpful conversations with J. K. Gamble (Sandia National Laboratories) and C. Lupien about the computer-assisted visual approach.  We acknowledge useful comments and discussions with G. Brookes, J. O. Simoneau and R. H. Foote.  This work was supported by the Natural Sciences and Engineering Research Council of Canada (NSERC) and the Canada Foundation for Innovation (CFI). This research was undertaken thanks in part to funding from the Canada First Research Excellence Fund.

All developed code and additional data are available upon request to \textit{maxime.lapointe-major@usherbrooke.ca}
\end{acknowledgments}

\bibliography{mabiblio}

\end{document}